\documentclass[lnbip]{svmultln}
\usepackage{makeidx}  
\usepackage{booktabs}
\usepackage{multirow}
\usepackage{graphicx}
\usepackage{tcolorbox}
\usepackage{array}
\usepackage{todonotes}
\usepackage{listings} 

\newcolumntype{C}[1]{>{\centering\arraybackslash}p{#1}}

\newcommand{\mypar}[1]{\smallskip\vspace{0.05cm}\noindent\textbf{#1.}}
\newcommand{\myparprompt}[1]{\smallskip\vspace{0.003cm}\noindent\textbf{#1}\vspace{-0.2cm}}

\tcbset{
  myboxstyle/.style={
    colback=gray!5, 
    colframe=gray!80!black, 
    fonttitle=\bfseries, 
    boxsep=2.3pt, 
    left=2.3pt, 
    right=2.3pt, 
    top=2.3pt, 
    bottom=2.3pt,
    before skip=5.0pt,
    after skip=3.0pt
  }
}

\usepackage{draftwatermark} 
\SetWatermarkText{Accepted Manuscript} 
\SetWatermarkScale{3.36}
\SetWatermarkAngle{50}
\SetWatermarkColor{gray!33}

\begin{document}
\raggedbottom
\sloppy
\mainmatter              
\title{ExOAR: Expert-Guided Object and \\Activity Recognition from Textual Data\thanks{Accepted manuscript on August 22, 2025, to the 2nd International Workshop on Generative AI for Process Mining (GenAI4PM 2025), held in conjunction with the 7th International Conference on Process Mining (ICPM 2025).}}
\titlerunning{Expert-Guided Object and Activity Recognition}  
%
\author{Iris Beerepoot\inst{1} \and Vinicius Stein Dani\inst{1} \and
Xixi Lu\inst{1}}
\authorrunning{Iris Beerepoot et al.}   
%
%
\institute{Utrecht University, Utrecht, The Netherlands,\\
\email{i.m.beerepoot@uu.nl}}

\maketitle              

\begin{abstract}        
Object-centric process mining requires structured data, but extracting it from unstructured text remains a challenge. We introduce ExOAR (Expert-Guided Object and Activity Recognition), an interactive method that combines large language models (LLMs) with human verification to identify objects and activities from textual data. ExOAR guides users through consecutive stages in which an LLM generates candidate object types, activities, and object instances based on contextual input, such as a user's profession, and textual data. Users review and refine these suggestions before proceeding to the next stage. Implemented as a practical tool, ExOAR is initially validated through a demonstration and then evaluated with real-world Active Window Tracking data from five users. Our results show that ExOAR can effectively bridge the gap between unstructured textual data and the structured log with clear semantics needed for object-centric process analysis, while it maintains flexibility and human oversight. 
\end{abstract}
\section{Introduction}
The availability of unstructured textual data from user interactions, such as window titles, document names, and application metadata, offers new opportunities for process analysis~\cite{Beerepoot2023,Rojas2024}. However, converting this data into structured formats suitable for object-centric process mining remains a significant challenge. A key challenge lies in identifying the objects and activities embedded in~the~text.

This challenge is evident across a variety of real-world domains. In healthcare, clinical case management depends on free-text sources like physician notes and discharge summaries, where extracting meaningful information about patients, treatments, and procedures requires expert knowledge due to domain-specific language and context~\cite{Munoz2022}. In IT service management, helpdesk systems generate large volumes of tickets and chat logs containing informal and evolving terminology, making it difficult to consistently identify issue types and resolution steps~\cite{Gupta2020}. In academic settings, digital traces like window tracking logs or file names encode complex, personal workflows that only the individual involved can meaningfully interpret~\cite{Beerepoot2023}. Similarly, in industrial manufacturing, maintenance logs and shift reports rely on technical shorthand and equipment codes known only to local staff~\cite{Pas2024}. These examples underscore a shared limitation: fully automated extraction methods often fall short due to ambiguity or domain specificity, while fully manual approaches are too labor-intensive to scale.

Recent progress in large language models (LLMs) has enabled new methods for semantic interpretation~\cite{Neuberger2025,Maxim2023,Yang2024}. While LLMs can suggest meaningful labels and associations, their outputs are often inconsistent or insensitive to domain-specific nuances~\cite{SteinDani2025}. At the same time, fully manual labeling remains time-consuming and error-prone~\cite{SteinDani2023Supporing}.

We introduce ExOAR (Expert-Guided Object and Activity Recognition), an interactive method that integrates the use of an LLM with expert validation. Users provide contextual input and the LLM is prompted to generate suggestion for candidate object types, activities, and objects. These suggestions are then reviewed and refined by the user and then used to enrich events, enabling object-centric process mining. ExOAR’s novelty lies in its human-in-the-loop design, making it especially effective where automated extraction alone falls short. 
We implement ExOAR as a practical tool and demonstrate its use on textual data from one of the authors. While the case study focuses on a specific data source, the approach is data-agnostic and extensible. A preliminary evaluation with another two academics and two professionals from outside of academia further illustrates its usability and effectiveness. 

\section{Background and Related Work}
\label{sec:background}

This section outlines key concepts in object-centric process mining and the challenges of extracting structured event data from unstructured sources. We also review recent efforts using LLMs for semantic recognition.

\subsection{Object-Centric Process Mining}

Object-centric process mining (OCPM) has emerged as a set of techniques that analyze processes in complex, multi-entity scenarios. Unlike traditional process mining techniques, OCPM allows for the mining of processes involving multiple object types and their interactions. Each event can be associated with one or more objects of different types, capturing the rich relational structure in many real-world processes~\cite{Aalst2019}. A prerequisite for object-centric process mining is the availability of a structured, object-centric event log (OCEL), in which the events are explicitly linked to relevant object types and instances. However, in many domains, especially those characterized by unstructured or semi-structured textual data, such an OCEL is not readily available. Deriving this detailed information about objects, object types, and activities from unstructured data sources, such as clinical notes, support tickets, or maintenance logs, is significantly more demanding than extracting a traditional event log~\cite{Guangming2018}. The unstructured nature of such data introduces new challenges, such as ambiguous references to objects, contextual or temporal dependencies between actions, domain-specific object types and terminology, and implicit relationships. 

Consequently, advanced log extraction techniques are essential to bridge this semantic gap between unstructured or semi-structured textual data and the structured OCELs required for OCPM~\cite{Rebmann2022}.
In this work, we approach this semantic gap by leveraging LLMs within a \mbox{human-in-the-loop}~design.

\subsection{Case or Object Recognition}
In conventional XES event log-based process mining, a case represents a single instance of a process, such as a customer order, a patient treatment cycle, or a support ticket, whose events are grouped by a unique case identifier. The case notion is foundational to discovering control-flow models and analyzing execution behavior. However, many real-life systems are not process-aware (e.g., a GPS tracking system, a web traffic log, or industrial devices) and the recorded events lack explicit case identifiers, a problem referred to as \textit{elusive cases}~\cite{suriadi2017event}.

Existing approaches to case recognition assume a single case notion and rely on heuristics, rules~\cite{Bayomie2022}, or auxiliary data to reconstruct case identifiers~\cite{Burattin2011,zetzsche2025case}. Others use process semantics and optimization approaches~\cite{BayomieCRM19}. However, these methods still misclassify or entirely discard events, especially when only partial information is available or when cases overlap.
More importantly, they assume the main object type is known and used as the case notion. To build an object-centric event log from unstructured or semi-structured textual sources, the object types and the number of object types may well be unknown. Additionally, the object identifiers of different types are also unknown and may even be of different formats. 
For example, in an academic setting, relevant object types may be courses, research projects, conference names, student names, etc.; and for each type the identifiers may have a different format, e.g., course names or even course codes, research paper titles, grant proposal titles, conference names, etc. 
Identifying relevant object types, their object instances, and correctly associating events with these instances becomes particularly challenging.

Recent advances in LLMs present a novel opportunity: LLMs can extract semantic-aware entities from free text~\cite{dagdelen2024structured}, enabling more adaptive and context-aware case recognition. However, their outputs can be inconsistent, especially in domain-specific contexts~\cite{SteinDani2025}. This underscores the need for expert-in-the-loop approaches, combining LLM-generated suggestions with domain knowledge to iteratively refine object types, object instances, and \mbox{activity assignments}.

\subsection{Semantic Recognition Using LLMs}
Previous studies have explored activity recognition from natural language text. 
In~\cite{Rebmann2023}, the authors presented a method for identifying and classifying user activities from low-level interaction data in real-time, without prior training or labeled examples. However, it focused on user behavior within applications, not on interpreting semantic content from textual artifacts such as window titles or document names. As such, it does not address the extraction of latent object and activity information from highly variable and domain-specific unstructured text. In turn, Leopold et al.~\cite{Leopold2011} proposed a method for generating meaningful names for process models by analyzing the activities and events within them. Their work presupposes that process elements are already labeled and organized within formal modeling notations. Consequently, their approach does not consider the extraction of those activities or objects from raw, unstructured textual sources or the integration of human expertise in ambiguous contexts.

Other works are using LLMs for various other semantic recognition tasks. In~\cite{Schneider2024}, LLMs were used to translate natural language into structured queries for question answering over knowledge graphs. 
Guo et al.~\cite{Guo2023} introduced a method to enhance wireless communication by using LLMs to rank words by semantic importance, allocating more transmission power to key words to reduce the chance of communication errors.
In the context of system-generated logs, Sani et al.~\cite{DBLP:conf/icpm/SaniSB23} and Shirali et al.~\cite{DBLP:conf/bpm/ShiraliSAS24} explore the use of LLMs to preprocess complex, fine-grained system-level logs, from user interfaces (UI logs) and Internet of Things (IoT) devices.
Sani et al.~\cite{DBLP:conf/icpm/SaniSB23} apply LLMs in RPA contexts to support task grouping, labeling, and connector recommendation, transforming low-level UI interactions into meaningful activities. 
Shirali et al.~\cite{DBLP:conf/bpm/ShiraliSAS24} propose prompt-based methods to abstract and integrate sensor-level events from IoT-sourced logs into activities.
However, both studies assume a single-case perspective with known cases, while our work addresses a more complex scenario of events linked to multiple, unknown entities (object types and objects).

Finally, Buss et al.~\cite{DBLP:conf/caise/BussKKRSW25} proposed a two-step LLM-based pipeline to extract object-centric event logs from text, consisting of a collector and a refiner. While their approach effectively identifies and cleanses semantic information from natural language, it does not delineate multiple interpretation steps within the collector phase, and their evaluation is limited to synthetic texts.

In this work, we build on these developments by leveraging LLMs' semantic recognition capabilities to support the identification of object types, objects, and activities to improve LLM-assisted object-centric event log generation from real-life textual data extracted from systems.

\section{Approach}
\label{sec:approach}
We propose ExOAR as a modular approach for object and activity recognition from textual data. The approach is designed for settings in which data contains implicit references to process-relevant entities and actions, but where human interpretation remains crucial for extracting meaningful information. While LLMs can support recognition by generating candidate object types, activities, and object instances from textual input, they are prone to ambiguity, inconsistency, and overgeneralization when used without context or oversight. The ExOAR approach addresses this by interleaving LLM-driven recognition with concise user validation in an iterative cycle. 

\begin{figure}[!b]
    \centering
    \includegraphics[width=\linewidth]{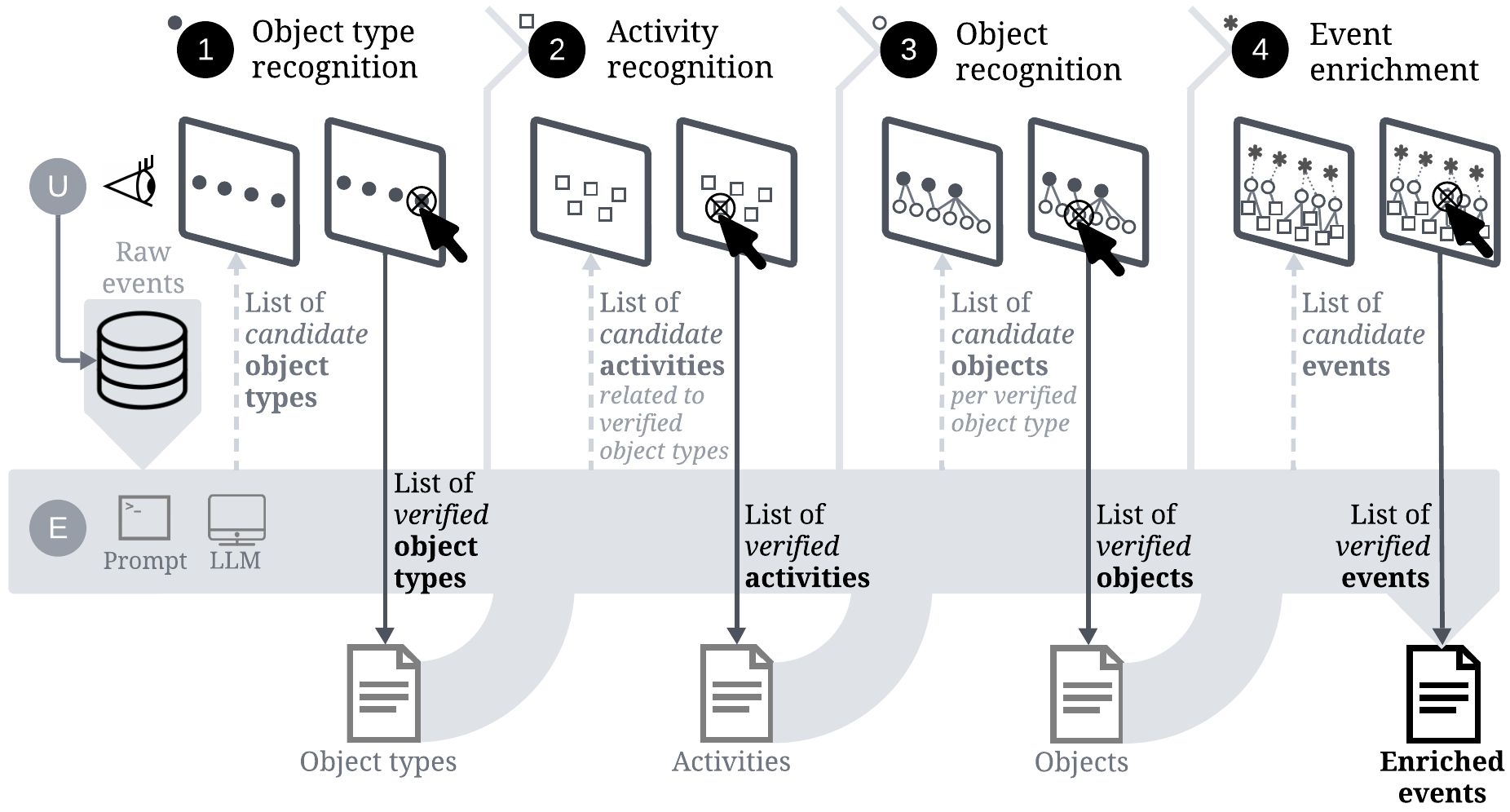}
    \caption{Overview of our approach, depicting a User (U) and ExOAR (E). }
    \label{fig:approach-overview}
\end{figure}

The approach focuses on three objectives: 
(1)~minimizing the effort required from end users, (2) ensuring economical use of LLM queries, and (3)~\mbox{offering} modularity in both the process and the types of data used. It decomposes the object and activity recognition process into four iterative steps (cf., Figure~\ref{fig:approach-overview}), where the ultimate goal is to enrich events from a textual data source for the purpose of enabling object-centric process mining. Next, we describe each step of our approach. 

\mypar{Step 1: Object type recognition} The first step focuses on identifying relevant \textit{object types}, i.e., general categories of entities that participate in a process. The user begins by supplying contextual knowledge that allows the LLM to interpret the textual data. This input is used to prompt the LLM to generate a list of candidate object types that are likely to appear in the user’s work context, after which resulting list is presented to the user for review. The user can edit this list by adding or removing object types. The output of this step is a verified set of object types that serve as anchors for subsequent recognition steps.

\mypar{Step 2: Activity recognition} In the second step, the focus shifts to recognizing relevant \textit{activities}, i.e., the types of actions users perform in relation to the identified object types. In this step, the information from the previous step is provided, i.e., the contextual knowledge and previously verified object types. This input is used to build a prompt for the LLM, which returns a list of candidate activity labels that are semantically linked to the provided context. The resulting list is then reviewed by the user, who can edit it as needed, adding or removing activities, confirming activities that most accurately reflect the types of work captured in the data. 
The result of this step is a curated set of activity labels that together with the object types form the input for the next steps.

\mypar{Step 3: Object recognition} The third step identifies concrete object instances, that is, specific entities that appear in the data and instantiate one of the previously verified object types. The prompt includes the provided data and verified object types and activities, allowing the model to align candidate objects with the semantic structure established in earlier steps. The LLM-generated output consists of a list of object instances, each assigned a likely object type. The user reviews and edits this list. 
The verified object instances provide the final ingredient for the last step in our approach.

\mypar{Step 4: Event enrichment} The final step enriches the textual data containing events with semantic annotations by associating texts with specific objects and activities identified in the previous steps. This step aims to approximate structured event-object-activity tuples that can serve as a foundation for object-centric process mining. The LLM is provided with the dataset as well as the verified object types, its instantiated objects, and activity labels. The LLM analyzes the semantic content of each event and proposes corresponding object and activity associations. Titles that are ambiguous may be left unannotated. From this enriched set, a sample of candidate annotations (i.e., a list of events with associated tuples of objects and activities) is presented to the user for validation and refinement. 
This final verification ensures high-quality associations are retained. The resulting dataset combines human judgment and LLM output, enabling a traceable, semantically grounded representation of the process behavior contained in the textual dataset.

\section{Demonstration}
\label{sec:demonstration}
We demonstrate our approach using one of the authors' Active Window Tracking (AWT) data \cite{Beerepoot2023}, before evaluating it with four other potential end users. An app was developed in Streamlit\footnote{Streamlit at: https://streamlit.io/} to support the process of recognizing objects and activities from AWT data\footnote{App homepage link, screenshots, and more information at: https://edu.nl/nhyjj}. Streamlit is an open-source Python framework to create interactive web-applications. We used OpenAI's gpt-4.1 model for the generation of output, because it offers a good balance between cost of usage and reliable combination of reasoning, coherent context understanding, and precise instruction-following\footnote{OpenAI documentation: https://openai.com/index/gpt-4-1/}. However, this model can be swapped with any other model of choice. Full details about the walkthrough, used dataset, intermediate results, and initial OCEL, are available online\footnote{Demonstration walkthrough at: https://edu.nl/pnyh9}. The walkthrough cost us \$0.08 in OpenAI credits. 
On the app's homepage, the user enters their profession, OpenAI API key and uploads (a subset of) their AWT data. After that, it is possible to continue to the first step. For the purpose of this walkthrough, we entered ``\textit{Academic staff}'' as profession on the homepage and uploaded the author's data for the full month of April 2025. 
Next, we describe each step of the walkthrough.

\begin{figure}[!h]
    \centering
    \includegraphics[width=\linewidth, trim=0cm 1.26cm 0cm 0cm, clip]{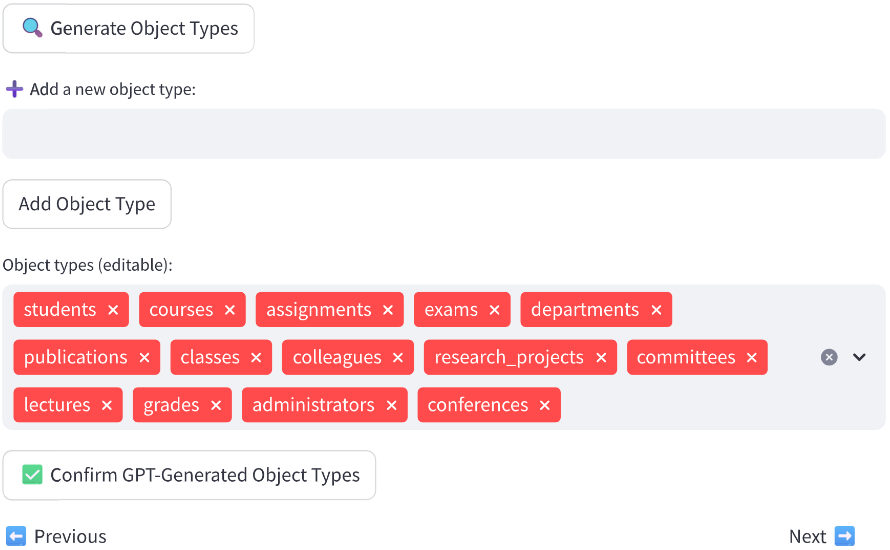}
    \caption{Screenshot of Step 1 - Object Identification}
    \label{fig:step1}
\end{figure}

\mypar{Step 1: Object type recognition} In this step, the user's profession is provided as input. The prompt is shown in the colorbox below (output format and example not shown for brevity). The output, i.e., the generated object types for the profession ``\textit{Academic staff}'', contained a list of 13 candidate object types. The tool allows us to add or delete object types from the list before continuing to \textit{Step~2}. We added ``\textit{conferences}'', as a significant portion of the author's time is spent on tasks related to academic conferences. We removed ``\textit{classes}'', ``\textit{grades}'', and ``\textit{administrators}'', as those are not core aspects of the author's work. A screenshot is provided in Figure \ref{fig:step1}; additional screenshots of the other steps and the complete results can be found in the repository.

\begin{tcolorbox}[title=Prompt Step 1, myboxstyle]
\scriptsize
You are an assistant specialized in semantic object recognition. Your task is to identify high-level object types based on a user’s profession. Object types represent general categories, human and non-human, and are used in object-centric event logs to group related entities.

\myparprompt{Task}
\begin{enumerate}
  \item Analyze the provided profession and reflect on the types of entities commonly involved in that work.
  \item Identify a list of relevant object types that could occur in the user's work processes.
  \item Focus on categories of entities, not specific instances or activities.
\end{enumerate}
\vspace{-0.2cm}

\myparprompt{Guidelines}
\begin{itemize}
  \item Output only lowercase string literals.
  \item Do not include any explanation or commentary.
  \item Include a broad range of object types that are reasonably relevant to the profession.
  \item Ensure object types are distinct and profession-relevant.
\end{itemize}

\end{tcolorbox}

\mypar{Step 2: Activity recognition} In this step, the profession  and confirmed object types are provided to the LLM as input. The output, i.e., the generated candidate activities, contained 20 items. Examples of these are ``\textit{prepare lectures}", ``\textit{supervise students}", and ``\textit{collaborate with colleagues}". We removed none, because they were all considered relevant, and added four more activities that we also considered relevant but were not in the initial set of candidate activities: ``\textit{attend department meetings}", ``\textit{prepare for committee meetings}", ``\textit{develop assignments}", and ``\textit{analyze research data}". This resulted in a total of 24 \mbox{confirmed activities.}

\begin{tcolorbox}[title=Prompt Step 2, myboxstyle]
\scriptsize
You are an assistant specialized in semantic activity recognition. Your task is to identify high-level work activities based on a user's profession and relevant object types. Activities describe meaningful steps a user performs and often reflect actions in business processes.

\myparprompt{Task}
\begin{enumerate}
  \item Analyze the provided profession and reflect on the types of activities commonly involved in that work.
  \item Analyze the provided object types. For each, identify a set of activities that typically involve, affect, or support those object types.
  \item Generate a list of relevant activities for the provided profession and object types. Include both core and supportive tasks that are recognizable within the user’s domain.
\end{enumerate}

\vspace{-0.2cm}
\myparprompt{Guidelines}
\begin{itemize}
  \item Return a list of lowercase strings.
  \item Do not include explanations or additional formatting.
  \item Focus on meaningful, commonly performed activities suitable for process mining.
\end{itemize}

\end{tcolorbox}

\mypar{Step 3: Object recognition} In the third step, the profession and object types are again provided as input, along with the set of activities confirmed in the previous step, as well as a set of active window titles. For practical and economic reasons, we provided the 500 most frequent window titles that appeared on 3 or more days for identifying relevant objects. The reasoning behind this was that this set should contain the objects that are most relevant to the person. The output contained 58 candidate object and object type combinations. We changed the title of the object in~10 instances to make the object name more precise, and edited the associated object types 6 times (e.g., one student was classified as colleague). We discarded 17 suggestions that were duplicates of other objects or irrelevant (e.g., the author's name was included under colleagues). This left us with~39 confirmed objects: 7 colleagues, 2 students, 3 courses, 2~assignments, 1 exam, 5~publications, 10 conferences, 2~committees, 1~department, and 6~research~projects. For example, the objects ``\textit{Stein Dani, V. (Vinicius)}" and ``\textit{Lu, X. (Xixi)}" were stored as colleagues, and ``\textit{ICPM 2025 (7th International Conference on Process Mining)}" and ``\textit{CoopIS 2024}" as conferences. 

\begin{tcolorbox}[title=Prompt Step 3, myboxstyle]
\scriptsize
You are an assistant specialized in extracting object instances from textual digital traces.  
Your task is to identify distinct object instances and assign them to appropriate object types.  
This is part of preparing structured data for object-centric process mining.

\myparprompt{Task}
\begin{enumerate}
  \item Analyze a list of window titles in context of a given profession, confirmed object types, and activities.
  \item Identify specific objects (e.g., "project alpha", "thesis john doe") mentioned or implied in those titles.
  \item Assign each object to the most appropriate type from the provided list.
\end{enumerate}

\vspace{-0.2cm}
\myparprompt{Guidelines}
\begin{itemize}
  \item Do not repeat objects (no duplicates).
  \item Do not assign the object name to be exactly the same as its object type.
  \item If the object type is a person (e.g., student, colleague), use a plausible name as object.
  \item Consider abbreviations, concatenations, or project/document references in titles.
  \item The result should help map interactions to real-world entities.
\end{itemize}

\end{tcolorbox}

\mypar{Step 4: Event enrichment} In the final step, the profession, selected object types, activities, and objects are all provided as input, along with a set of~100 frequent window titles. Providing a set of 100 titles is again a practical and economical reason, as the aim was to arrive at a set of 10 titles that can be verified by the user. In reviewing the sample, we found that all 10 titles were correctly given labels, although we did make changes to some activities and objects. We kept four events as-is, but edited the activities in four other instances. Specifically, we added ``\textit{attend department meetings}'' to one as that was also a relevant activity, added ``\textit{analyze research data}'' in another, and changed ``\textit{design exams}'' to ``\textit{grade exams}'' in a third as it was considered more fitting. Finally, we deleted ``\textit{organize conferences}'' in the last one, as the conference title did appear in the window title, but the activity was not related to organizing the conference. 
We also added an additional object to two titles. Tables~\ref{tab:activity_counts} and \ref{tab:object_counts} illustrate how often each activity and object occur in the reviewed sample. 

\begin{tcolorbox}[title=Prompt Step 4, myboxstyle]
\scriptsize
You are an assistant specialized in associating textual titles with objects and activities relevant to professional workflows.  
Your task is to infer meaningful semantic associations between window titles and known entities.

\myparprompt{Task} \vspace{0.2cm}\\
For each of the following window titles, determine whether it clearly relates to one or more of the given activities and one or more of the given objects.  
If so, return the title and its associated activities and objects. Otherwise, return only the title with empty lists.

\myparprompt{Guidelines}
\begin{itemize}
  \item Use your understanding of the user's profession to ground your associations.
  \item Include objects and activities only if they are directly and unambiguously implied.
  \item Avoid guessing or over-interpreting vague titles.
\end{itemize}

\end{tcolorbox}

\vspace{-0.3cm}
\begin{table}[!h]
\centering
\footnotesize

\begin{minipage}[t]{0.42\textwidth}
\centering
\caption{Frequency of activities in reviewed sample}
\label{tab:activity_counts}
\resizebox{0.8\textwidth}{!}{%
\begin{tabular}{lc}
\hline
\textbf{Activity} & \textbf{Count} \\
\hline
collaborate with colleagues & 3 \\
attend department meetings & 1 \\
manage research projects & 2 \\
analyze research data & 1 \\
design exams & 1 \\
grade exams & 2 \\
attend conferences & 1 \\
present at conferences & 1 \\
coordinate with departments & 1 \\
participate in committees & 1 \\
review publications & 1 \\
\hline
\end{tabular}
}
\end{minipage}
\hfill
\begin{minipage}[t]{0.4\textwidth}
\centering
\caption{Frequency of objects in reviewed sample}
\label{tab:object_counts}
\resizebox{0.808\textwidth}{!}{%
\begin{tabular}{lc}
\hline
\textbf{Object} & \textbf{Count} \\
\hline
Reijers, H.A. (Hajo) & 1 \\
Department of ... (ICS) & 2 \\
AWT2OCEL & 1 \\
Lu, X. (Xixi) & 1 \\
Stein Dani, V. (Vinicius) & 1 \\
BPM Exam Remindo & 2 \\
Interstellar Dagstuhl team & 1 \\
Interstellar paper & 1 \\
Process Mining Camp 2025 & 1 \\
AI Lab & 1 \\
WI25 & 1 \\
\hline
\end{tabular}
}
\end{minipage}

\end{table}

\vspace{-0.3cm}
\section{Evaluation}
\label{sec:evaluation}
We performed a preliminary evaluation with two additional academic staff members and two users from other professions, namely a bookkeeper and a self-employed business advisor. Each participant had collected Active Window Tracking data for at least a month, and uploaded their own data in the tool. We guided the users through the evaluation and discussed the results of each step. 

Table \ref{tab:evaluation-summary} shows the results of the evaluation\footnote{Object types and activities can be found here: https://edu.nl/wcjpg}. For the academics, 11 object types were generated and kept as-is. However, Academic A added \textit{``reviews''} as an additional object type. For the bookkeeper, 14 object types were confirmed, whereas for the self-employed business advisor, 17 object types were selected. When it comes to the activities, the tool generated 24 activities for the academic staff profession, which were all retained by the academics. Academic A added \textit{``hire staff''} and \textit{``develop software''}, while Academic B added \textit{``attend project meetings''} and \textit{``advise employees''}. 

\begin{table}[!bt]
\footnotesize
\centering
\caption{Summary of evaluation results}
\label{tab:evaluation-summary}
\resizebox{0.921\textwidth}{!}{%
\begin{tabular}{@{}p{2.1cm}p{2.1cm}p{2.7cm}p{2.7cm}p{2.7cm}p{3.0cm}@{}}
\toprule
\textbf{Step} 
& \textbf{Metric} 
& \textbf{Academic A} 
& \textbf{Academic B} 
& \textbf{Bookkeeper} 
& \textbf{Business Advisor} \\
\midrule
\multirow{4}{=}{\textbf{Object Types}} 
    & Generated         & 11 & 11 & 14 & 15 \\
    & Kept as-is        & 11 (100\%) & 11 (100\%) & 13 (93\%) & 15 (100\%)\\
    & Added             & 1 & 0 & 1  & 2\\
    & Removed           & 0 & 0 & 1  & 0\\
\midrule
\multirow{4}{=}{\textbf{Activities}}   
    & Generated         & 24 & 24 & 15 & 20\\
    & Kept as-is        & 24 (100\%) & 24 (100\%) & 15 (100\%)  & 18 (90\%)\\
    & Added             & 2 & 2 & 0  & 0\\
    & Removed           & 0 & 0 & 0  & 2\\
\midrule
\multirow{4}{=}{\textbf{Objects}}      
    & Generated         & 60 & 50 & 83 & 55\\
    & Kept as-is        & 42 (70\%) & 48 (96\%) & 56 (67\%) & 41 (75\%)\\
    & Edited            & 5 & 0 & 8  & 9\\
    & Removed           & 18 & 2 & 19 & 5\\
\midrule
\multirow{4}{=}{\textbf{Events} (verified sample = 10)}       
    & Generated         & 23 & 18 & 28 & 52\\
    & Kept as-is        & 3 (30\%) & 5 (50\%) & 8 (80\%)  & 7 (70\%)\\
    & Edited            & 7 & 4 & 1  & 3\\
    & Removed           & 0 & 1 & 1  & 0\\
\bottomrule
\end{tabular}
}
\end{table}
 
Thus, for the object types and activities, minimal corrections were needed. When it comes to the objects and events, there is more variety in the accuracy of results. A significant number of objects was discarded across participants, because they were irrelevant (e.g., they referred to the user themselves as employee, or their own firm as a client) or they were considered duplicates of another object. The object identification step also triggered a reflection on the earlier selection of object types. In our current set-up, adding an additional object type in step~1 required re-running of the generation in subsequent steps, which we wanted to avoid for economic purposes and making the best use of our evaluation participants' time. However, from this discussion, we conclude that users (even the academics familiar with object-centric process mining) may find it challenging to identify relevant object types a priori. A solution may be to suggest a more diverse set of object types, since it is easier for users to discard irrelevant suggestions than to conceive of missing ones. False negatives can limit the usefulness of the output, particularly because subsequent steps rely on the initial \mbox{set of object~types.} 

As for the final object and activity associations to events, this was assessed positively. Only two events were removed across the four participants, as the titles were too generic or vague to confidently associate activities and objects. The edits that were made mostly related to deleting one or more activities from the generated list. For example, when the window titles of the academics included a reference to an academic conference, it would list \textit{``attend conferences"}, \textit{``present at conferences"}, \textit{``organize conferences"} all as corresponding activities. In several instances, the academics would only keep \textit{``organize conferences"}. 

In future versions of the tool, we might consider changing the prompt to explicitly ask for the best-fitting activities, although this might also lead to the exclusion of relevant activities. As noted earlier, it is easier to discard irrelevant suggestions than conceive of missing ones, so perhaps erring on the side of overgeneration is the best \mbox{solution as of now}. 

\section{Conclusion}
\label{sec:conclusion}
In this paper, we introduced ExOAR, a novel expert-guided approach for object and activity recognition from textual data to support object-centric process mining. By combining large language models with human validation across four modular steps, ExOAR effectively bridges the gap between unstructured textual traces and the structured event-object-activity representations required for analysis. Our demonstration and preliminary evaluation across diverse professions highlight the tool’s flexibility, usability, and potential to generalize across domains. While results indicate strong performance in identifying object types and activities with minimal corrections, challenges remain in refining object instances and ensuring consistent event enrichment. Future work will focus on scaling evaluations, improving prompt strategies, and enhancing guidance for object type selection to further increase robustness and adoption. Ultimately, we envision ExOAR enabling large-scale semantic interpretation of workplace data, opening possibilities for transparent, personalized digital productivity support in complex domains such as healthcare, education, or legal services.

\vspace{0.3em}\noindent\textbf{Acknowledgment}
We are grateful for the users who participated in our evaluation. ChatGPT assisted with code generation and  language refinement.

%
%
\bibliographystyle{splncs04} 
\bibliography{mybibliography}
\end{document}